\RequirePackage{lineno}
\documentclass[showpacs,aps,amsmath,amssymb,prc,superscriptaddress,floatfix,preprint,preprintnumbers]{revtex4}%,
\usepackage{graphicx}% Include figure files
\usepackage{dcolumn}% Align table columns on decimal point
\usepackage{bm}% bold math
\begin{document}
%\setkeys{acs}{articletitle = true}
%\preprint{APS/123-QED}

\title{Elliptical flow coalescence to identify the $f_{0}$(980) content}% Force line breaks with \\

\author{An Gu}%4ed with \\
\email{gu180@purdue.edu}
\affiliation{Department of Physics and Astronomy, Purdue University, West Lafayette, IN 47907, USA}
\author{Terrence Edmonds}
\email{tedmonds@purdue.edu}
\affiliation{Department of Physics and Astronomy, Purdue University, West Lafayette, IN 47907, USA}
\author{Jie Zhao}
\email{zhao656@purdue.edu}
\affiliation{Department of Physics and Astronomy, Purdue University, West Lafayette, IN 47907, USA}
\author{Fuqaing Wang}
\email{fqwang@purdue.edu}
\affiliation{Department of Physics and Astronomy, Purdue University, West Lafayette, IN 47907, USA}
\affiliation{School of Science, Huzhou University, Huzhou, Zhejiang 313000, China}
\date{\today}% It is always \today, today,
% but any date may be explicitly specified

\begin{abstract}
We use a simple coalescence model to generate $f_{0}$(980) particles for three configurations: a ${s\bar{s}}$ meson, a ${u\bar{u}s\bar{s}}$ tetraquark, and a ${K^{+}K^{-}}$ molecule. The phase-space information of the coalescing constituents is taken from a multi-phase transport (AMPT) simulation of heavy-ion collisions. It is shown that the number of constituent quarks scaling of the elliptic flow anisotropy can be used to discern ${s\bar{s}}$ from ${u\bar{u}s\bar{s}}$ and ${K^{+}K^{-}}$ configurations.
\end{abstract}


%display desiredgenerate $f_{0}$(980) particles for three configurations: ${s\bar{s}}$ meson, ${u\bar{u}s\bar{s}}$ tetraquark
\maketitle

%======================================
%==============================================
%==============================================
%=============================
\section{Introduction}

Exotic hadrons (hadrons with configurations other than the usual ${q\bar{q}}$ and ${qqq(\bar{q}\bar{q}\bar{q})}$ configurations) have been searched for a long time, since exotic hadron states are allowed by quantum chromodynamics (QCD) and therefore their studies can further our understanding of QCD \cite{Jaffe:2004ph}. The $f_{0}$(980) is one of the candidate exotic hadrons which was first observed in ${\pi\pi}$ scattering experiments in the 1970's \cite{ Protopopescu:1973sh, Hyams:1973zf, Grayer:1974cr}. Its configuration is still controversial--- it can be a normal ${s\bar{s}}$ meson, a tetraquark ${s\bar{s}q\bar{q}}$ state, or a ${K\bar{K}}$ molecule \cite{Bugg:2004xu, Klempt:2007cp, Pelaez:2015qba}.

Heavy ion collisions create a deconfined state of quarks and gluons, called the quark-gluon plasma (QGP) \cite{Adcox:2004mh,Arsene:2004fa, Back:2004je, Adams:2005dq, Muller:2012zq}. They can provide a suitable environment to study exotic hadrons, because a large number of quarks and gluons permeate the QGP. When the temperature decreases, those quarks and gluons group into hadrons, presumably including exotic ones. The process is called hadronization process and is not well understood. A common mechanism to describe hadronization in heavy-ion collisions is the quark coalescence in which several quarks(antiquarks) combine together to form a hadron \cite{Dover:1991zn,Fries:2008hs}. Coalescence model was originally developed to describe the formation of deutrons from targets exposed to proton beams \cite{Butler:1963pp} and is extensively used to describe hadron production in relativistic heavy ion collisions \cite{Dover:1991zn,Fries:2003vb,Fries:2003kq,Greco:2003mm,Minissale:2015zwa}.

In non-central heavy ion collisions, the azimuthal distribution of particles is anisotropic, believed to result from hydrodynamic expansion of the initial anisotropic overlap regions \cite{Ollitrault:1992bk}. The particle azimuthal distribution is often expressed in Fourier series \cite{Voloshin:1994mz}:

\begin{eqnarray}\label{equ:flows}
\frac{\mathrm{d}N}{\mathrm{d}\phi}\propto1+2\sum_{n=1}^{\infty}v_{n}\cos[n(\phi-\psi_{n})],
\end{eqnarray}
where $\phi$ is the particle azimuthal angle, $\psi_{n}$ is the $n$-th harmonic plane. The coefficients ($v_{n}$) are often called anisotropic flows, and are transverse momentum ($p_{T}$) and rapidity ($y$) dependent. In heavy-ion collisions, the leading anisotropic term is the $n=2$ term because of the approximate elliptical shape of the collision overlap geometry; $\psi_2$ is a proxy for the unmeasured reaction plane and $v_{2}$ is called elliptic flow. If partons (quarks, antiquarks) which combine into a hadron have the same momentum, then we have
\begin{eqnarray}\label{equ:pth}
p_{T,h}=n_{q}\cdot p_{T,q}, 
\end{eqnarray}
where $n_{q}$ is the number of constituent quarks in the hadron. Keeping only $v_{2}$ in Eq.( \ref{equ:flows}), we have
\begin{eqnarray}
\frac{\mathrm{d}N_{h}}{\mathrm{d}\phi}\propto\left(\frac{\mathrm{d}N_{q}}{\mathrm{d}\phi}\right)^{n_q}\propto\left[1+2v_{2,q}(p_{T,q})\cos(2[\phi-\psi_{RP}])\right]^{n_q}\\\nonumber
\approx1+n_q \cdot 2v_{2,q}(p_{T,q})\cos(2[\phi-\psi_{RP}]).
\end{eqnarray}
Thus, we have
\begin{eqnarray}
v_{2,h}(p_{T,h})=n_{q}\cdot v_{2,q}(p_{T,h}/n_q). \label{equ:v2scaling}
\end{eqnarray}
This result is known as the number of constituent quarks (NCQ) scaling of elliptic flow, when the momenta of the coalescing (anti)quarks are not identical, the NCQ scaling is not as good \cite{Molnar:2003ff}. Approximate NCQ scaling has been observed experimentally \cite{Adams:2003am, Adams:2004bi, Adams:2005zg, Adare:2006ti, Abelev:2007rw, Adamczyk:2013gw,Abelev:2014pua}. The elliptic flow of a hadron species can therefore tell us the number of constituent quarks contained in the hadron.

In this work, we use a coalescence model to study the elliptic flow ($v_{2}$) of the $f_{0}$(980) for its different configuration assumptions. 
Although the string melting version of the AMPT model (a multiphase transport) \cite{Lin:2004en} uses quark coalescence to form hadrons \cite{Lin:2003jy,He:2017tla}, it does not produce tetraquark hadrons. In order to simulate the production of the $f_{0}$(980) for different configurations, we build our own simple coalescence model. We take the phase-space information of quarks (and Kaons) from AMPT in mid-central Au+Au collisions at $200A$ GeV as input to our coalescence. We use this simple coalescence model to generate pions, protons, Kaons, $\phi$ mesons, and $f_{0}$(980) particles of three configurations (${s\bar{s}}$, ${u\bar{u}s\bar{s}}$, ${K\bar{K}}$) and calculate their elliptic flow. We first compare the $v_{2}$ of pions and protons from our coalescence model with those from AMPT to validate our simple coalescence model approach. We then study the NCQ scaling of the $f_{0}$(980) $v_{2}$ and demonstrate that it is a viable way to identify its quark content.

%=============================
\section{Coalescence Model}
The main idea of the coalescence model is to combine several partons into one hadron. The model was implemented in heavy ion collisions to describe the NCQ scaling of elliptic flow, the baryon-to-meson ratio, and the hadron transverse momentum spectra, which can not be described well by fragmentation model \cite{Fries:2003kq}. 

Suppose N constituent particles are coalesced into a composite particle (a hadron or a ${K\bar{K}}$ molecule). The total yield of the composite particle can be expressed as \cite{Dover:1991zn}

\begin{eqnarray}
{N_{c}}=g_{c}\int \left( \prod_{i=1}^{N} \mathrm{d}N_{i} \right) f_{c}^{W}(\vec{r_{1}},\cdots, \vec{r_{N}},\vec{p_{1}},\cdots, \vec{p_{N}}) \label{equ:coalescence1}.
\end{eqnarray}
Here $f_{c}^{W}(\vec{r_{1}},\cdots, \vec{r_{N}},\vec{p_{1}},\cdots, \vec{p_{N}})$ is the Wigner function (WF) which is proportional to the coalescence probability and $g_{c}$ is a statistical factor. The statistical factor $g_c$ only affects the yield of the hadrons instead of the elliptic flow, so we set $g_c=1$ for all kinds of hadrons in this study.
%common used wigner function

%harmonic oscillator wigner funciton

%${v_{2}}$ number of constituent quarks scaling

For a meson, if two quarks form a harmonic oscillator and they are in s-state, then the WF is Gaussian \cite{Baltz:1995tv}:

\begin{eqnarray}
{f_{meson}(\vec{r_{1}},\vec{r_{2}},\vec{p_{1}},\vec{p_{2}})}=&&
A \cdot \exp \left(-\frac{r_{12}^{2}}{{\sigma_{r}}^{2}}-\frac{p_{12}^{2}}{{\sigma_{p}}^{2}}\right)\label{equ:mesona},
\end{eqnarray}
where
\begin{eqnarray}
{r_{ij}^{2}}=(\vec{r_{i}}-\vec{r_{j}})^{2}\label{equ:mesonb}, \qquad {p_{ij}^{2}}=(\vec{p_{i}}-\vec{p_{j}})^{2}.\label{equ:mesonc}
\end{eqnarray}
Here, ${\vec{r_{i}}}$ and ${\vec{p_{i}}}$ are the position and momentum of i-th quark/antiquark at the time the hadron is formed. In AMPT, partons freeze out (FO, which means this (anti)quark doesn't interact with others anymore) at different times ${t_{i}}$. The moment for two or more (anti)quarks to coalesce is set to be the latest freeze out time of those (anti)quarks, ${t_{F}}$. The final positions are calculated as ${\vec{r}_{i}={\vec{r}}_{i,FO}+\vec{v}_{i,FO}\times(t_{F}-t_{i,FO})}$. For ${K}$ and ${\bar{K}}$ particles, we take their FO phase space information right after they are formed in AMPT.

There are two parameters in Eq.(\ref{equ:mesona}), ${A}$ and ${\sigma_{r}}$ ($\sigma_{p}=1/\sigma_{r}$). The parameter ${A\le 1}$ only affects the total yield of hadrons, not the ${v_{2}}$, so we set it to 1. Assuming the system is in s-wave state, we have $\sigma_{r}=1/\sqrt{\mu\omega}$, where $\mu$ is the reduced mass of the two-body system, and $\omega$ is the oscillator frequency. The oscillator frequency can be fixed by $\omega=3/(2\mu_{1}\langle r^{2} \rangle) $, where ${\langle r^{2} \rangle}$ is the mean square radius of the hadron \cite{Baltz:1995tv}. Thus, we have $\sigma_{r}=\sqrt{2\langle r^{2} \rangle/3 }$. For pions, ${\langle r^{2} \rangle=(0.61\pm0.15)\textnormal{ fm}^{2}}$ \cite{ADYLOV1974402}, so we set ${\sigma_{r}}=0.64\textnormal{ fm}$. For kaons, ${\langle r^{2} \rangle=(0.34\pm0.05)\textnormal{ fm}^{2}}$ \cite{Amendolia:1986ui}, we set ${\sigma_{r}}=0.48\textnormal{ fm}$. For phi mesons ($s\bar{s}$), its internal structure and radius is still not well known, and its cross section with nonstrange hadron is small \cite{Shor:1984ui}, so we set ${\sigma_{r}=0.5\textnormal{ fm}}$.

For multi-particle systems, the quantum state is difficult to compute analytically. For tetraquark and pentaquark hadrons, only the heavy quark sector has been calculated quantum mechanically using perturbative approaches \cite{Ali:2017jda,Chen:2016qju,PhysRevD.57.6778,Shi:2013rga}. In this work, for baryons and tetraquark systems, we naively define the Wigner function also to be Gaussian. For baryons, 
\begin{eqnarray}
{f_{baryon}(\vec{r_{1}},\vec{r_{2}},\vec{r_{3}},\vec{p_{1}},\vec{p_{2}},\vec{p_{3}})}=&&
A \cdot \exp \left( -\frac{1}{3 \sigma_{r}^{2}} \cdot{\sum\limits_{i,j=1 \ i < j}^{3} {r}_{ij}^{2}}-\frac{1}{3 \sigma_{p}^{2}}\cdot {\sum\limits_{i,j=1 \ i < j}^{3} {p}_{ij}^{2}} \right), \label{eq2a}
\end{eqnarray}
and for tetraquarks,
\begin{eqnarray}
{f_{tetra}(\vec{r_{1}},\vec{r_{2}},\vec{r_{3}},\vec{r_{4}},\vec{p_{1}},\vec{p_{2}},\vec{p_{3}},\vec{p_{4}})}=
&& A \cdot \exp \left( -\frac{1}{6 \sigma_{r}^{2}}\cdot {\sum\limits_{i,j=1 \ i < j}^{4} {r}_{ij}^{2}}-\frac{1}{6 \sigma_{p}^{2}} \cdot{\sum\limits_{i,j=1 \ i < j}^{4} {p}_{ij}^{2}} \right). \label{eq3a}
\end{eqnarray}
For protons, ${\sqrt{\langle r^{2} \rangle}=0.88\,\textnormal{fm}}$ \cite{1742-6596-312-3-032002}, so we set ${\sigma_{r}}=\sqrt{2\langle r^{2} \rangle/3 }=0.72\,\textnormal{fm}$.
For the $f_{0}$(980) particles, we have $\omega=67.8\,\textnormal{MeV}$ \cite{Cho:2010db}. If we consider $s\bar{s}$ configuration, the reduced mass is $\mu=m_{s}/2$ with $m_s=0.199\,\textnormal{GeV}/c^{2}$ \cite{Lin:2004en}, so $\sigma_{r}=1/\sqrt{\mu\omega}=2.4\,\textnormal{fm}$. We set this value of $\sigma_{r}$ for all three different configurations of the $f_{0}$(980).
In our coalescence model, we get the freeze out information of (anti)quarks after the parton cascade in AMPT. We input this information to our simple coalescence model to produce hadrons. We loop over all available (anti)quarks to form pions, protons, or $f_{0}$(980), and we carry out the coalescence separately for each of these species. For each species, if the flavors of the (anti)quarks are correct for the hadron and the value of a random number (uniformly distributed between 0 and 1) is smaller than the value of the Wigner function, the hadron is formed. The four momentum of the hadron is calculated as the sum of the four momentum of its constituents, ${p_{h}^{\mu}=\sum_{i}p_{q,i}^{\mu}}$. And these (anti)quarks are then removed from further consideration of coalescence. 

%=================================================
%=================================================
\section{Results}
We use this simple coalescence model to generate pions, protons, Kaons, $\phi$ mesons, and $f_{0}(980)$ of three different configurations (${s\bar{s}}$, ${u\bar{u}s\bar{s}}$, ${K^{+}K^{-}}$).

In each event, the elliptic flow ${v_{2}}$ \cite{Poskanzer:1998yz, Xiao:2012uw} is calculated as: 
\begin{eqnarray}
v_{2}=\left< \mathrm{\cos}{2(\phi-\psi_{2}^{(r)})} \right>.
\end{eqnarray}
Here ${\psi_{2}^{(r)}}$ is the 2nd harmonic plane of each event in the spatial configuration space of the initial overlap geometry, and is obtained by 
\begin{eqnarray}
\psi_{2}^{(r)}=\left[ \mathrm{atan2}(\left< r_{\perp}^{2} \mathrm{sin}2\phi_{r} \right>,\left<r_{\perp}^{2} \mathrm{\cos}2\phi_{r} \right>)+\pi \right]/2,
\end{eqnarray}
where ${r_{\perp}}$ and ${\phi_{r}}$ are the polar coordinate of each initial parton before the parton cascade \cite{Ollitrault:1993ba}. The resolution of $\psi_{2}^{(r)}$ is close to 1 due to the large initial parton multiplicity \cite{Xiao:2012uw}. The elliptic flow shown in this study are for particles within pseudo-rapidity window $|\eta|<1$. 

We first compare the results of pions and protons from our coalescence model with those from AMPT. We then present the ${f_{0}}$(980) results from our coalescence model.

%=============================
\subsection{Proton and Pion}
\begin{figure}
\includegraphics[height=6cm]{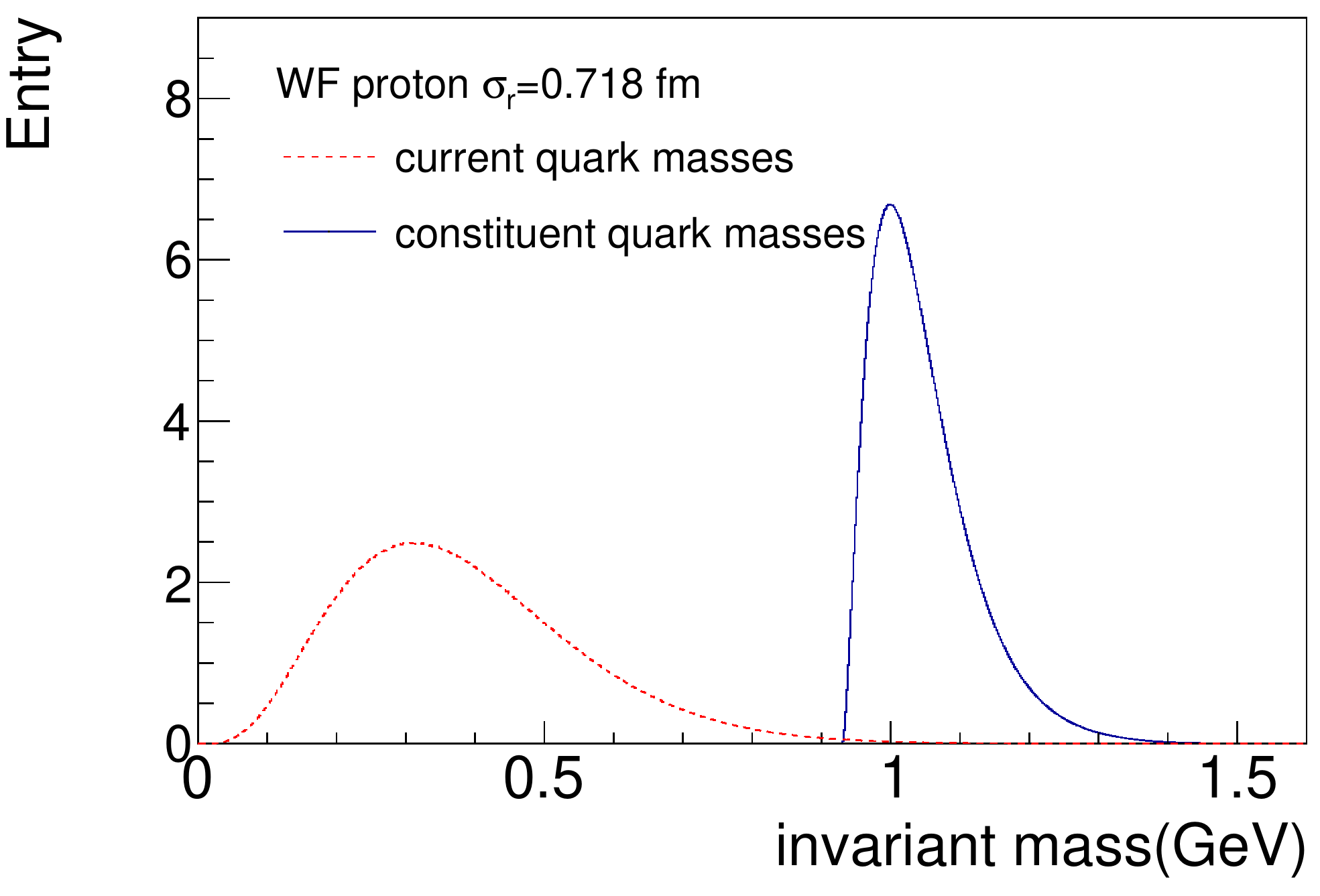}% Here is how to import EPS art
\caption{\label{fig:fig1} Invariant mass spectra of protons from our Gaussian-WF coalescence with current quark masses (as in AMPT) and constituent quark masses, respectively. }
\end{figure}

The quark masses used in our study are from the AMPT model (the same as PYTHIA program \cite{Lin:2004en}. It would be more reasonable to use the constituent quark masses \cite{BorkaJovanovic:2010yc} to take into account the effects of gluons, but as we show below, the quark masses do not significantly alter our results, so we stick to the masses used in the AMPT model.

Figure \ref{fig:fig1} shows the mass spectra of protons from our Gaussian-WF coalescence with AMPT quark masses (red line) and constituent quark masses (blue line) ($m_{u}=m_{d}=0.31 \,\textnormal{GeV}$ \cite{BorkaJovanovic:2010yc}). We can see that the coalescence model does not generate correct mass spectra, which is a well-known problem \cite{Lin:2004en, He:2017tla}. 

\begin{figure}
\includegraphics[height=6cm]{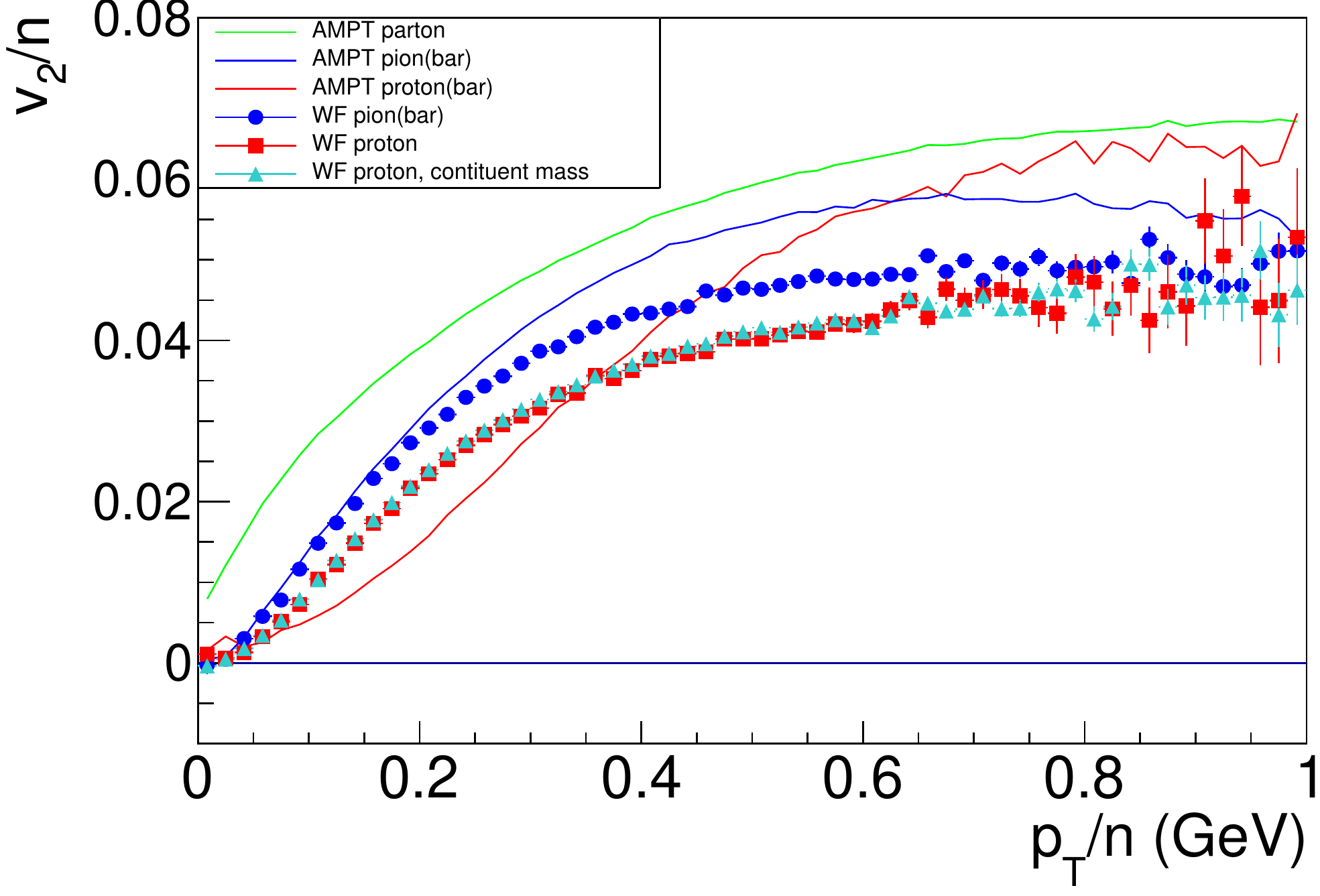}% Here is how to import EPS art
\caption{\label{fig:fig2}${v_{2}/n_q}$~vs.~${ p_{T}/n_q}$ of pions and protons from Gaussian-WF coalescence compared to those from AMPT.}
\end{figure}

Figure \ref{fig:fig2} shows ${v_{2}/n_q}$~vs.~${ p_{T}/n_q}$ for the pions (blue circle, $n_q=2$) and protons (red square, $n_q=3$) from our Gaussian-WF coalescence compared to those from AMPT model (blue line and red line). The pions from the Gaussian-WF coalescence have similar ${v_{2}/n_q}$ to pions from AMPT at low $p_T$ and lower ${v_{2}/n_q}$ than AMPT pions at high ${ p_{T}/n_q}$. While the WF protons have higher ${v_{2}/n_q}$ than AMPT protons at low $p_T$ but lower ${v_{2}/n_q}$ at high ${ p_{T}/n_q}$. 

We checked whether the masses of quarks would affect the ${v_{2}/n_q}$~vs.~${ p_{T}/n_q}$ by using constituent quark masses in coalescence. To do that, we simply take the AMPT quark freezeout momenta and recalculate their velocity using constituent quark masses in propagation of quarks from freeze out to coalescence point. The result is shown in Fig. \ref{fig:fig2} where ${v_{2}/n_q}$~vs.~${ p_{T}/n_q}$ of protons generated from our coalescence with constituent quark masses (light blue triangle) is also presented. The masses of quarks have practically no effect on the ${v_{2}/n_q}$. This is expected because in our simple coalescence model, the masses of quarks only affect the speeds of quarks (most relativistic), which are only related to the final spatial position of the quarks. While the ${v_{2}}$ only depends on the momentum of quarks, so the ${v_{2}}$ would be almost independent of the quark masses.

\begin{figure}
\includegraphics[height=6cm]{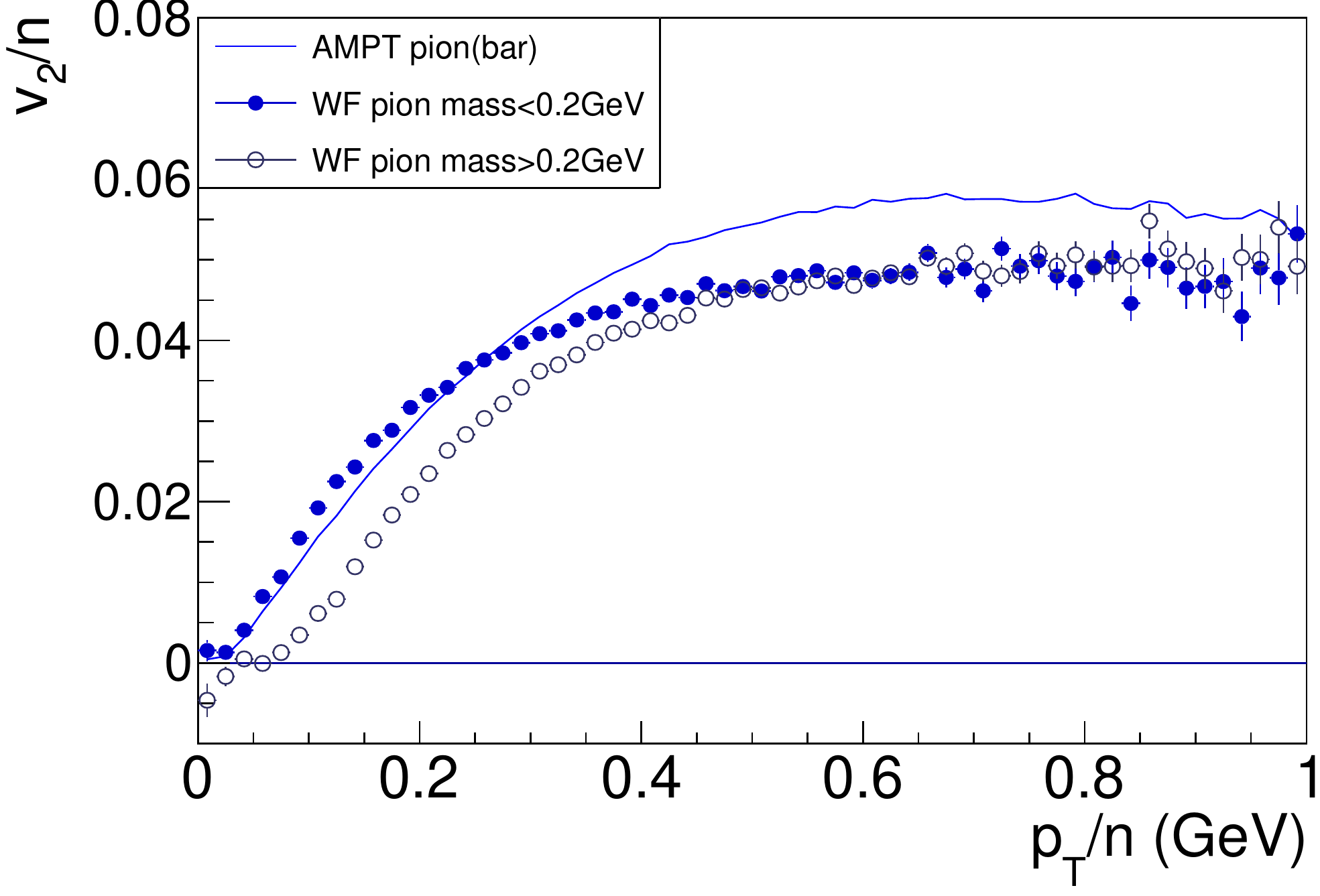}% Here is how to import EPS art

\includegraphics[height=6cm]{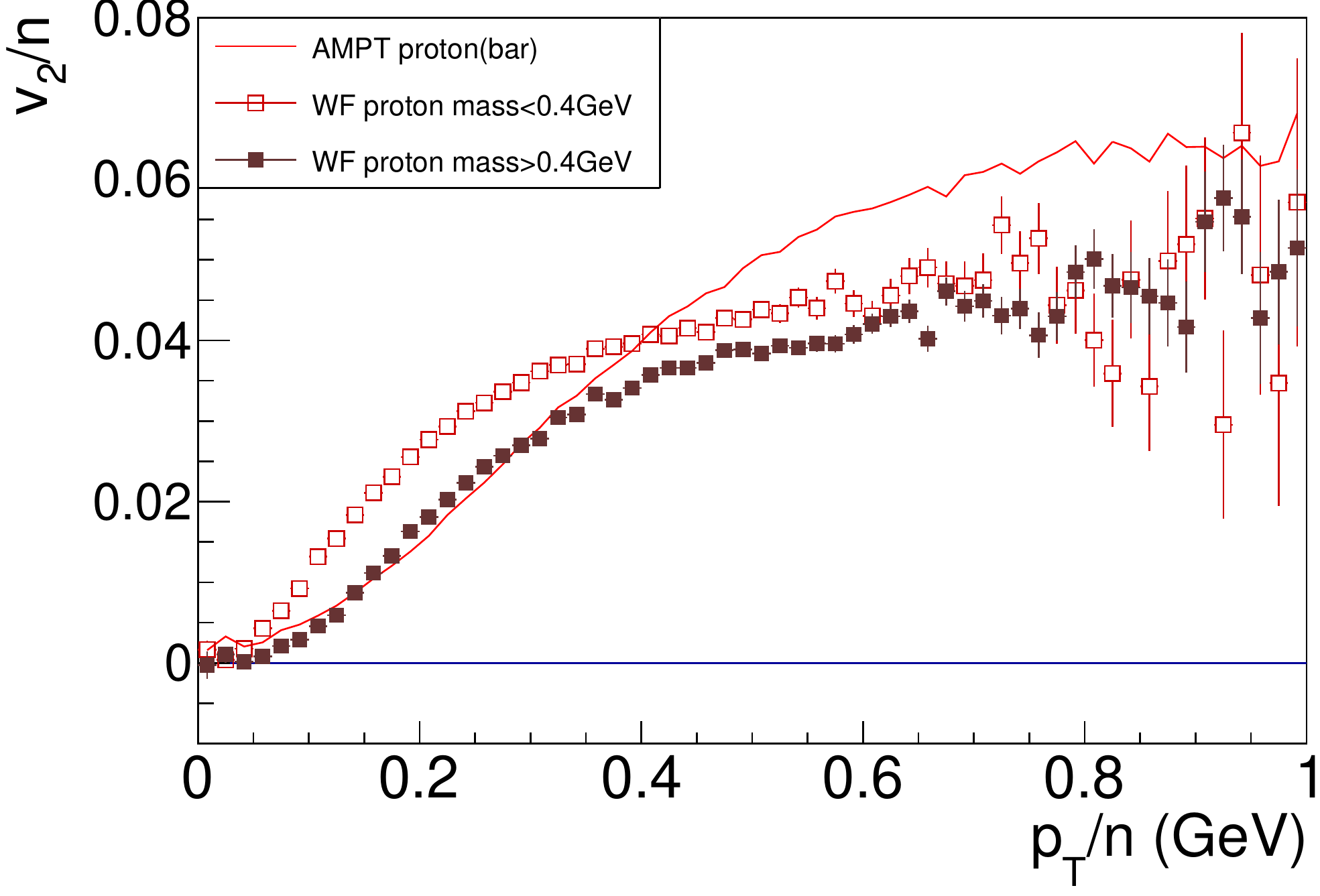}% Here is how to import EPS art
\caption{\label{fig:fig3}${v_{2}/n_q}$~vs.~${ p_{T}/n_q}$ of pions (upper) and protons (lower) from Gaussian-WF coalescence for different mass cuts, compared to AMPT results.
}

\end{figure}

We apply invariant mass cut on protons and pions from our Gaussian-WF coalescence as shown in Fig.\ref{fig:fig3}. Generally, hadrons with larger invariant masses have less ${v_{2}}$ since their constituent quarks are farther away from each other in momentum space. 
When mass${<}$0.2 GeV is applied to pions and mass${>}$0.4 GeV is applied to protons from the Gaussian-WF coalescence, our simple coalescence model gives more consistent results with those from the coalescence model used in the AMPT. This is understandable because the invariant mass is utilized by AMPT to assign hadrons \cite{Lin:2004en}. It is also an indication that our simple coalescence model is doing a reasonable job. 

\begin{figure}
\includegraphics[height=6cm]{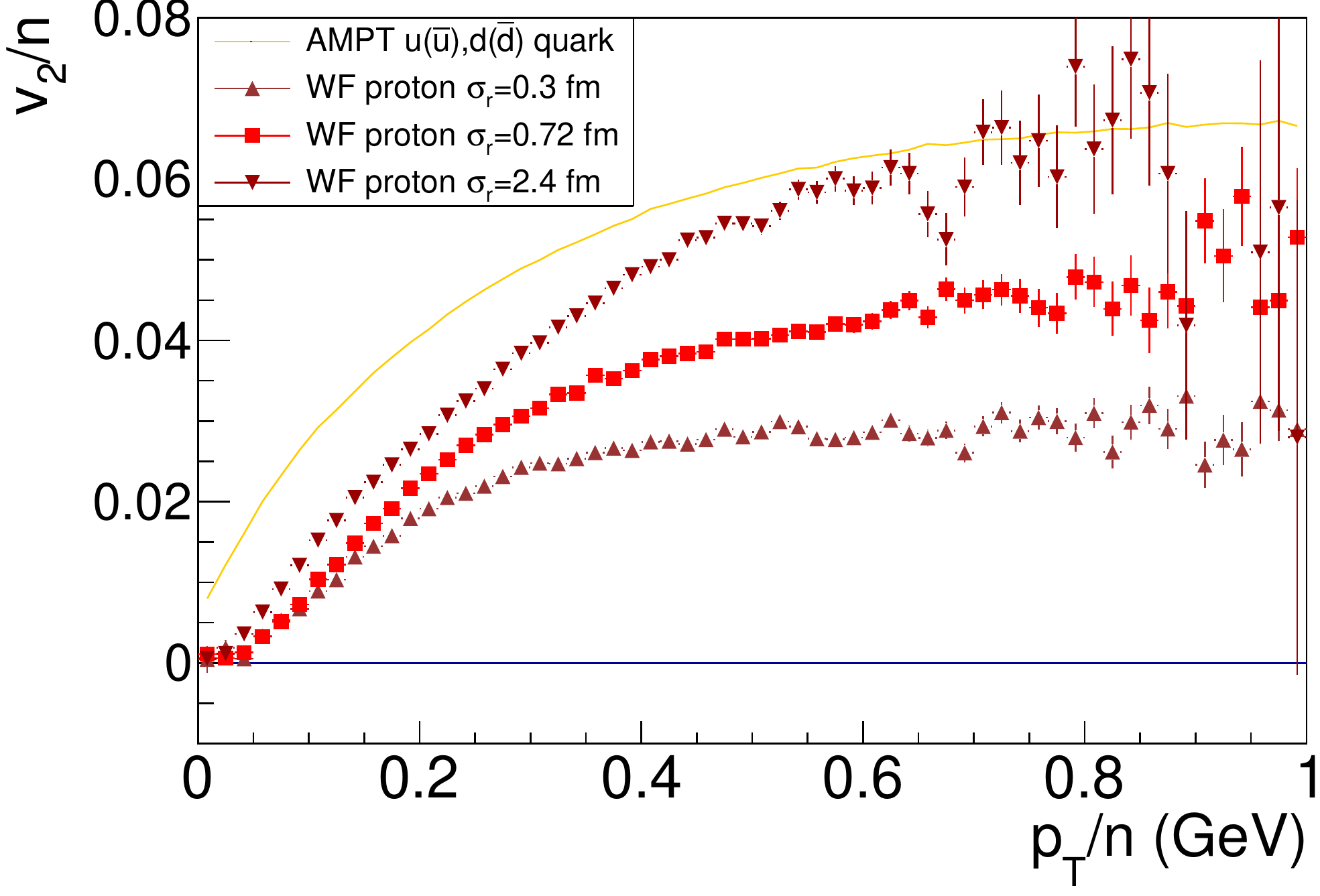}% Here is how to import EPS art
\caption{\label{fig:fig2_diff}${v_{2}/n_q}$~vs.~${ p_{T}/n_q}$ of protons generated with different values of $\sigma_r$ from our WF-Gaussian coalescence, compared to that of light quarks from AMPT.}
\end{figure}

As mentioned in the introduction, the NCQ scaling is best satisfied when the coalescing partons have the same momentum. When their momenta differ, the NCQ will not be as good. Since the smaller the $\sigma_r$ the larger their momentum difference ($\sigma_p=1/\sigma_r$), we expect the goodness of the NCQ scaling to increase with $\sigma_r$. Thus we artificially change the $\sigma_r$ of the proton and check how the proton ${v_{2}/n_q}$~vs.~${ p_{T}/n_q}$ changes relative to the light quark ${v_2}$. This is shown in Fig. \ref{fig:fig2_diff}. Indeed, we observe that protons with larger value of $\sigma_r$ have closer ${v_{2}/n_q}$~vs.~${ p_{T}/n_q}$ to that of the $u(\bar{u}), d(\bar{d})$ quarks, i.e. closer to the ideal situation of NCQ scaling of elliptic flow ($\sigma_p\rightarrow 0$). Likewise it is reasonable for the pions and protons to have different ${v_{2}/n_q}$~vs.~${ p_{T}/n_q}$ from the quarks as shown in Fig. \ref{fig:fig2} because of the relatively small values of $\sigma_r$ .

It is interesting to notice that although protons are generated with larger $\sigma_r$ ($\sigma_r=0.72$ fm) than the pions ($\sigma_r=0.64$ fm), the ${v_{2}/n_q}({ p_{T}/n_q})$ of protons is lower than that of pions. This is because there are more constituent quarks in a parton than a pion.  So there is a larger reduction in $v_2$ from the ideal NCQ scaling picture. That is to say, in our Gauss-WF coalescence model, hadrons containing more constituents will have lower $v_{2}/n_q(p_{T}/n_q)$ for a given limited value of $\sigma_r$.

% While the $f_{0}$(980) has almost the same ${v_{2}/n_q}$~vs.~${ p_{T}/n_q}$ as the $s(\bar{s}) as shown below in Fig. \ref{fig:fig

%=============================
\subsection{$f_{0}$(980) particle for three configurations(${s\bar{s}}$, ${u\bar{u}s\bar{s}}$, ${K^{+}K^{-}}$)}

\begin{figure}
\includegraphics[height=6cm]{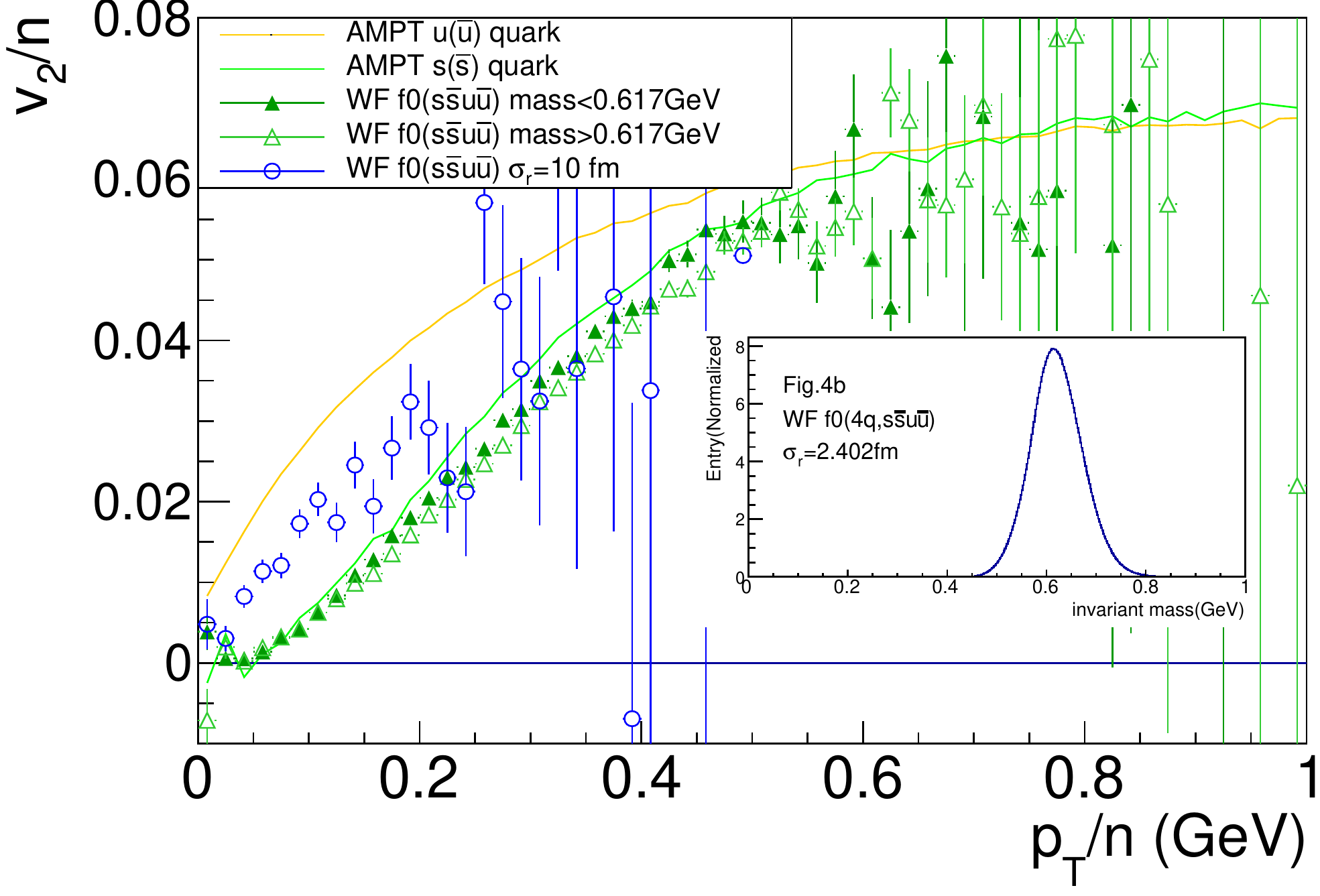}
\caption{\label{fig:fig5}${v_{2}/n_q}$~vs.~${ p_{T}/n_q}$, for the tetraquark state of $f_{0}$(980) from our Gaussian-WF coalescence ($\sigma_r=2.4$) fm for different mass cuts. For comparison, the $f_{0}$(980) results with $\sigma_r=10$ fm are also shown. The curves show the results of light and strange quarks from AMPT.
}
\end{figure}

Figure \ref{fig:fig5} shows ${v_{2}/n_q}$~vs.~${ p_{T}/n_q}$ ($n_q$=4) for the tetraquark state of $f_{0}$(980) from our Gaussian-WF coalescence for different mass cuts. The mass spectrum of $f_{0}$(980) in tetraquark state from our Gaussian-WF coalescence is shown in the insert. The mass spectrum of $f_{0}$(980) does not peak at 980 MeV. Using larger constituent quark masses could improve the situation. Similar to the protons shown in Fig.\ref{fig:fig3}, the mass cut does not significantly change the ${v_{2}/n_q}$ of $f_0(980)$ either. 

For comparison, the $v_2$ vs. $p_T$ of light quark and strange quark from AMPT are also shown in Fig.\ref{fig:fig5}. It is interesting to note that the $v_2/n_q(p_T/n_q)$ of the $f_0(980)$($u\bar{u}s\bar{s}$) is lower than those of u-quarks and s-quarks, whereas one would naively expect the the $f_0(980)$($u\bar{u}s\bar{s}$) results to be midway between u-quarks and s-quarks curves. In the ideal NCQ scaling picture, all the coalesced quarks in the hadron possess the same momentum, i.e. $\sigma_p\rightarrow 0$ and $\sigma_r\rightarrow\infty$. Hence the $v_{2}/n_{q}$ of $f_{0}(u\bar{u}s\bar{s})$ should be $(v_{2,u\bar{u}}+v_{2,s\bar{s}})/2$ because:

\begin{eqnarray}
\label{equ:usquark_v2}
\frac{\mathrm{d}N_{f_0(980)(u\bar{u}s\bar{s})}}{\mathrm{d}\phi}\propto\left[1+2v_{2,u}(p_{T,u})\cos(2[\phi-\psi_{RP}])\right]^{2}\cdot\left[1+2v_{2,s}(p_{T,s})\cos(2[\phi-\psi_{RP}])\right]^2\\\nonumber
\approx1+2\cdot\left[2v_{2,u}(p_{T,f_0(980)}/4)+2v_{2,s}(p_{T,f_0(980)}/4)\right]\cos(2[\phi-\psi_{RP}]).
\end{eqnarray}

\begin{figure}
\includegraphics[height=6cm]{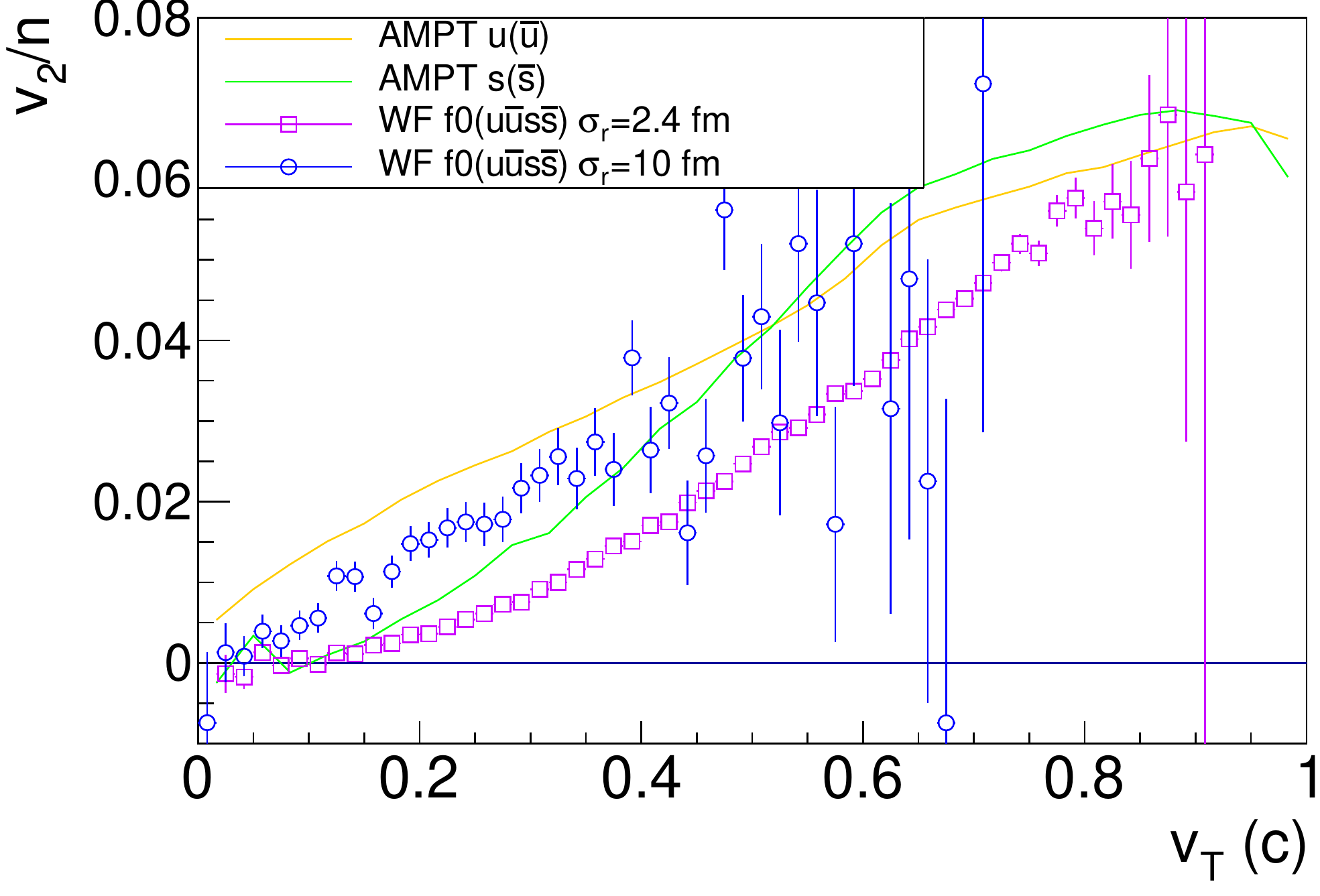}
\caption{\label{fig:fig10}${v_{2}/n_q}$~vs.~${ v_{T}}$, for the tetraquark state of $f_{0}$(980) from our Gaussian-WF coalescence for constituent quark masses and different values of $\sigma_r$. The curves are the results of up quark and strong quark from AMPT calculated by using constituent quark masses.
}
\end{figure}

To test this, we artificially set $\sigma_r$ to a large value ($\sigma_r=10$ fm) to mimic the ideal NCQ scaling picture. The results are shown in Fig.\ref{fig:fig5} as the blue open circles. Indeed, the $v_{2}(p_T/n_q)$ of the $f_{0}(u\bar{u}s\bar{s})$ lies midway between those of u-quarks and s-quarks as expected in the ideal picture.

In the coalescence, one often considers the velocities of the constituents. In order to examine the elliptic flow as a function of velocity, we use the large constituent quark masses ($m_u=0.31$ GeV, $m_s=0.5$ GeV) to produce the $f_0(980)(u\bar{u}s\bar{s})$. Similar to the protons, the $v_2/n_q(p_T/n_q)$ of $f_0(980)(u\bar{u}s\bar{s})$ using constituent quark masses is almost as same as that using constituent quark masses. The $v_2/n_q$ of $f_0(980)(u\bar{u}s\bar{s})$ as a function of transverse velocity $v_T$ is show in Fig.\ref{fig:fig10}  together with those of constituent u-quarks and s-quarks. When $\sigma_r$ of $f_0(980)(u\bar{u}s\bar{s})$ is set to 10 fm, the $v_2/n_q(v_T)$ also lies midway between those of u-quarks and s-quarks as expected in the ideal picture.  The $v_2/n_q(v_T)$ is qualitatively similar to the $v_2/n_q(p_T/n_q)$ shown in Fig.\ref{fig:fig5}. This is because the constituent masses of u-quarks and s-quarks are not very different. It should be noted, however, that in our Gauss-WF coalescence model, it is the momentum, not the velocity that is used to calculate the coalescence probability. Hence, the momentum is the more relevant variable to use than the velocity.

The above results indicate that our coalescence model is doing a reasonable job to produce tetraquark hadrons. In the following, we use our coalescence model to produce the $f_0(980)$ of the other two configurations ($s\bar{s}$ and $K\bar{K}$) and compare the elliptic flows of these three configurations.

\begin{figure}
\includegraphics[height=6cm]{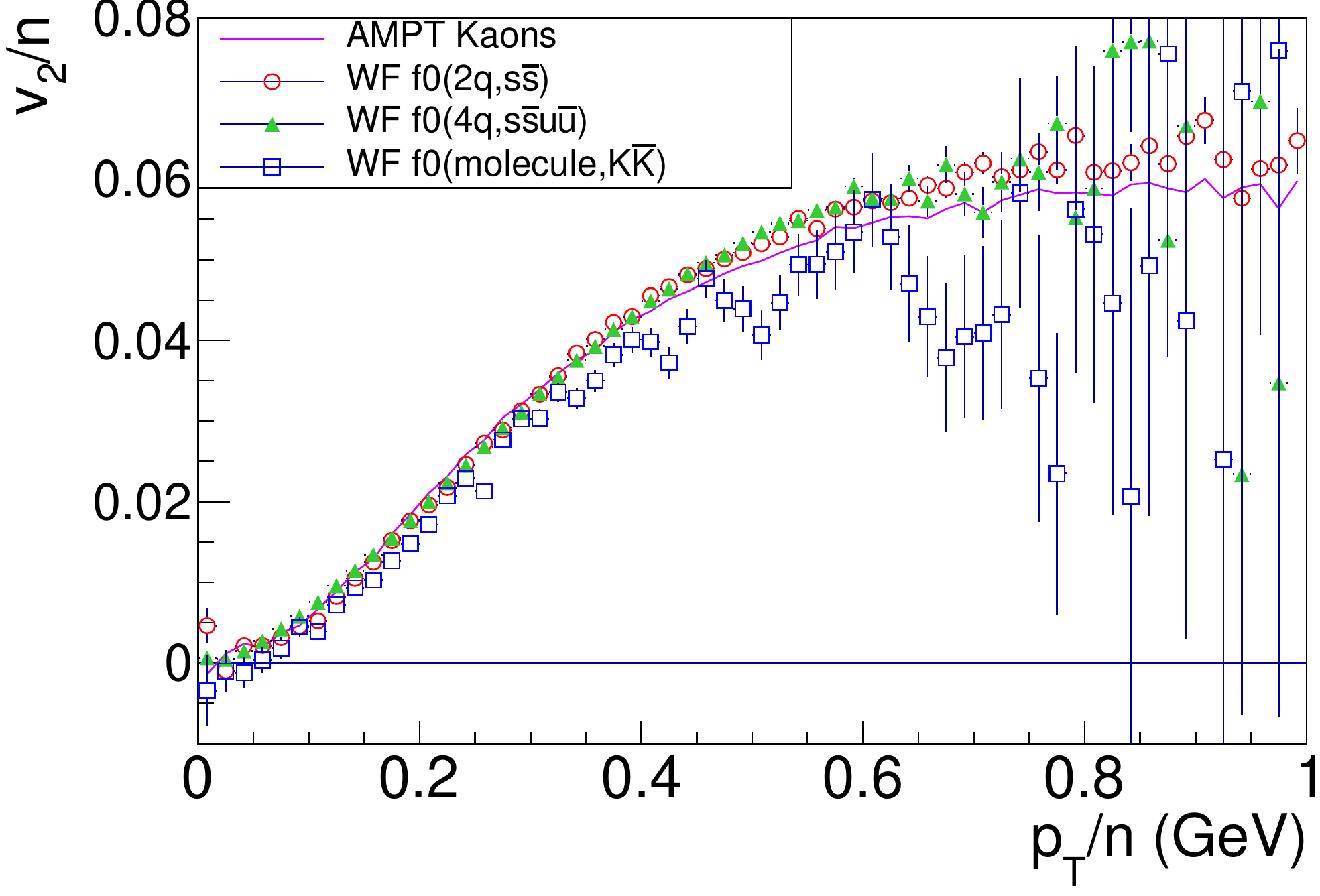}% Here is how to import EPS art
\caption{\label{fig:fig6}${v_{2}/n_q}$~vs.~${ p_{T}/n_q}$ for different configurations of ${f_{0}}$(980) from our Gaussian-WF coalescence compared to the Kaon result from AMPT.}
\end{figure}

In Fig.\ref{fig:fig6}, we compare $v_{2}/n_q$~vs.~$p_{T}/n_q$ of different configurations of $f_{0}$(980). The ${v_{2}/n_q}$ is almost the same for different configurations (${s\bar{s}}$, ${u\bar{u}s\bar{s}}$, ${K^{+}K^{-}}$). It is easy to understand why $f_{0}(980)(K^+K^-)$ ($n_q=4$) has the same $v_{2}/n_q(p_{T}/n_q)$ as the $f_{0}(980)(s\bar{s})$ ($n_q=2$). This is because they are both from the two-body coalescence with the same value of $\sigma_r$, and because in AMPT kaons ($n_q=2$) have almost the same $v_{2}/n_q(p_{T}/n_q)$ as that of the strange quarks ($n_q=1$). 

However, it is somewhat suprising that the $v_{2}/n_q(p_{T}/n_q)$ of $f_{0}(u\bar{u}s\bar{s})$ is almost the same as that of $f_{0}(s\bar{s})$.
As we have previously pointed out that in our Gauss-WF coalescence model, for a given value of $\sigma_r$, hadrons containing more constituents will have lower $v_{2}/n_q(p_{T}/n_q)$. Thus we would expect the $v_{2}/n_q(p_{T}/n_q)$ of $f_{0}(u\bar{u}s\bar{s})$ to be lower than that of the $f_{0}(s\bar{s})$, which is not true here. This is because of another effect, i.e.~the u-quarks that $f_{0}(u\bar{u}s\bar{s})$ contains have larger $v_{2}/n_q(p_{T}/n_q)$ and upraise that of the $f_{0}(u\bar{u}s\bar{s})$. As a result, the $f_{0}(980)$ of $u\bar{u}s\bar{s}$ and $s\bar{s}$ configurations happen to have almost the same $v_{2}/n_q(p_{T}/n_q)$.

\begin{figure}
\includegraphics[height=6cm]{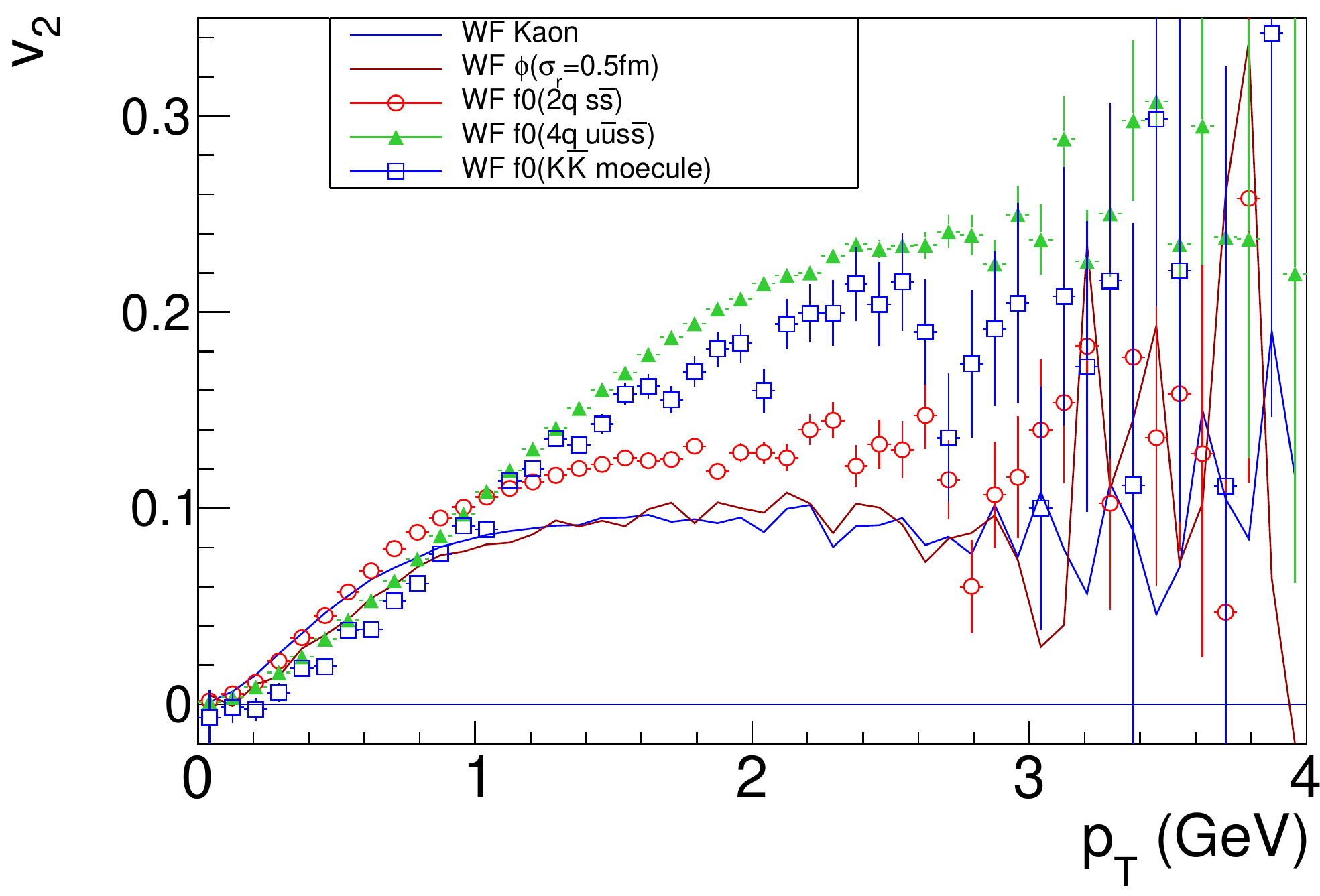}% Here is how to import EPS art
\caption{\label{fig:fig7}${v_{2}}$~vs.~${ p_{T}}$, for Kaons, $\phi$, and three different configurations of ${f_{0}}$(980) from our Gaussian-WF coalescence model.}
\end{figure}

The common dependence of the $v_2/n_q$ vs. $p_T/n_q$ in Fig.\ref{fig:fig6} indicates that the NCQ scaling of $f_0(980)$ can be used to tell its number of constituent quarks. This is more evidently shown in Fig.\ref{fig:fig7} where the $v_2$ is directly shown as a function of $p_T$. The $v_{2}(p_{T})$ of ${f_{0}}$(980) with (${s\bar{s}}$) configuration is very different from the other configurations, especially when ${ p_{T}>1}$ GeV. 
So according to our simple coalescence model, experimental measurement of ${v_{2}}$ can tell whether ${f_{0}}$ particle is composed of 2 quarks. 
It is however difficult to tell the difference between the 4-quark configuration and ${K^{+}K^{-}}$ molecule configuration. This is not surprising because the ${K^{+}K^{-}}$ molecule is effectively a ``four-quark" state.

Studying the yield of exotic hadrons in heavy ion collisions using coalescence model is another way to discriminate between different configurations. It is shown that the yield of an exotic hadron (such as a tetraquark state) is significantly smaller than the yield of a non-exotic hadron with normal number of constituent quarks \cite{Cho:2010db}. This can be used to further separate tetraquark ${f_{0}}$(980) from a ${K^{+}K^{-}}$ molecule state.
%==============================================
%==============================================
\section{Conclusion}
We used a simple coalescence model with Gaussian Wigner function to generate pions, protons, kaons, $\phi$ mesons, and $f_{0}(980)$ particles of three different configurations (${s\bar{s}}$, ${u\bar{u}s\bar{s}}$, ${K^{+}K^{-}}$). The NCQ scaling of elliptic flow $v_2$ is observed in our study, and can be used to distinguish the ${s\bar{s}}$ state of the ${f_{0}}$(980) from the tetraquark (${u\bar{u}s\bar{s}}$) or ${K^{+}K^{-}}$ molecule state in heavy ion collisions. It is difficult to tell apart the ${u\bar{u}s\bar{s}}$ and ${K^{+}K^{-}}$ states by measuring ${v_{2}}$. The ${f_{0}}$(980) yields needs to be exploited.
%==============================================
\section*{Acknowledgement}
We thank Dr.~Pengfei Zhuang and Dr.~Zi-wei Lin for fruitful discussions.
This work was supported in part by the U.S.~Department of Energy (Grant No.~de-sc0012910) and the National Natural Science Foundation of China (Grant No.~11747312).

\newpage %Just because of unusual number of tables stacked at end
\bibliographystyle{unsrt} %include paper titles
\bibliography{apssamp}% Produces the bibliography via BibTeX.

%
%\bibliography{ref} %include your ref's in ref.bib

%\bibitem{vidakovic1994wavelets}
\end{document}